# Beyond Contrast Transfer: Spectral SNR as a Dose-Aware Metric for STEM Phase Retrieval

Georgios Varnavides[1,2], Julie Marie Bekkevold[3], Stephanie M Ribet[2], Mary C Scott[2,4], Lewys Jones[3], Colin Ophus[5]

[1] Miller Institute for Basic Research in Science, University of California Berkeley, Berkeley, USA
[2] National Center for Electron Microscopy, Lawrence Berkeley National Laboratory, Berkeley, USA
[3] School of Physics, Trinity College Dublin, Dublin, Ireland
[4] Department of Materials Science and Engineering, University of California Berkeley, Berkeley, USA
[5] Department of Materials Science and Engineering, Stanford University, Stanford, USA

Correspondence may be addressed to: Georgios Varnavides, gvarnavides@berkeley.edu; Colin Ophus, cophus@gmail.com

## Abstract

The contrast transfer function (CTF) is widely used to evaluate phase retrieval methods in scanning transmission electron microscopy (STEM), including center-of-mass imaging, parallax imaging, direct ptychography, and iterative ptychography. However, the CTF reflects only the maximum usable signal, neglecting the effects of finite electron fluence and the Poisson-limited nature of detection. As a result, it can significantly overestimate practical performance, especially in low-dose regimes. Here, we employ the spectral signal-to-noise ratio (SSNR), as a dose-aware statistical framework to evaluate the recoverable signal as a function of spatial frequency. Using numerical reconstructions of white-noise objects, we show that center-of-mass, parallax, and direct ptychography exhibit dose-independent SSNRs, with close-form analytic expressions. In contrast, iterative ptychography exhibits a surprising dose dependence: at low fluence, its SSNR converges to that of direct ptychography; at high fluence, it saturates at a value consistent with the maximum detective quantum efficiency predicted by recent quantum Fisher information bounds. The results highlight the limitations of CTF-based evaluation and motivate SSNR as a more accurate, dose-aware metric for assessing STEM phase retrieval methods.

## Introduction

High-resolution imaging of radiation-sensitive samples remains a fundamental challenge in scanning transmission electron microscopy (STEM) (Egerton, 2013, 2007). Biological and soft materials, such as proteins and organic solids, scatter electrons weakly and are highly susceptible to radiation damage, particularly from inelastic interactions that drive radiolysis. As a result, characterizing such materials at high resolution with conventional imaging approaches is challenging (Liu et al., 2020), highlighting the need for dose-efficient imaging techniques with accurate information transfer. STEM phase retrieval methods, which attempt to reconstruct the sample potential from diffraction data, show promise for dose-efficient imaging (Bekkevold et al., 2024, Küçükoğlu et al., 2024, Li et al., 2025, Yu et al., 2024). However, evaluating their performance, in terms of spatial information transfer and dose efficiency, requires more accurate and dose-aware metrics.

The most common metric for evaluating STEM phase retrieval techniques is the contrast transfer function (CTF), which characterizes how well spatial frequency components of the sample are transmitted through an ideal linear imaging system (Bekkevold et al., 2025, Ma and Muller, 2024; Yang et al., 2016, 2015; Yu et al., 2024). While useful for optimizing parameters, the CTF represents only the maximum transmittable signal and does not account for finite electron fluence or the Poisson-limited nature of electron detection (O'Leary et al., 2021, Seki et al., 2018).

As a result, CTF-based assessments can overestimate the practical performance of imaging methods, especially in the low-dose regimes relevant for sensitive specimens. This discrepancy is most evident in iterative ptychography, where the CTF can be shown numerically to equal unity (Bekkevold et al., 2025, Ma and Muller, 2024, Varnavides et al., 2023), yet low-dose reconstructions of biological samples exhibit 'Thon-like' oscillations (see Figure 1, right half). To address this limitation, we employ the spectral signal-to-noise ratio (SSNR) as a dose-aware metric for evaluating information transfer and dose-efficiency in phase retrieval techniques (Unser et al., 1987).

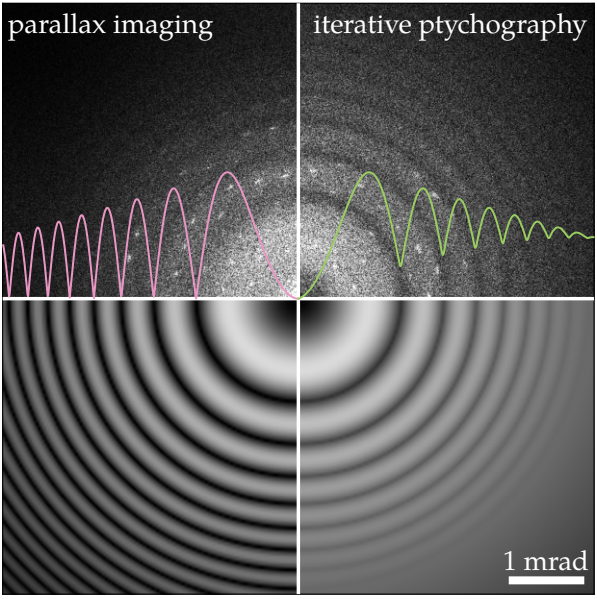

Figure 1: Power spectrum of experimental parallax imaging (top-left) and iterative ptychography (top-right) reconstructions of biological crystals (Küçükoğlu et al., 2024), exhibiting strong 'Thon-like' oscillations. Note that the iterative ptychography and parallax oscillations have different functional forms, with the latter exhibiting no zero crossings. The fitted CTFs (bottom half) and radially-averaged line plots are derived in the Spectral Signal-to-Noise Ratio section.

The manuscript is structured as follows. First, we briefly review the CTFs of commonly-employed phase retrieval techniques, following Hammel and Rose (1995) and Bekkevold et al. (2025). Specifically, we focus on center-of-mass imaging (Dekkers and De Lang, 1974, Lazić et al., 2016), parallax imaging (Spoth et al., 2017, Varnavides et al., 2024, Yu et al., 2024), single side-band (SSB) direct ptychography (Pennycook et al., 2015, Yang et al., 2015), and iterative ptychography (Chen et al., 2020, Rodenburg and Maiden, 2019). We interpret the geometric and mathematical origins of the probe autocorrelation envelope and show how the direct ptychography CTF is readily obtained as a single iteration of iterative ptychography.

We perform independent numerical reconstructions on white-noise objects, and quantify their statistics using the SSNR metric (Unser et al., 1987). For center-of-mass imaging, parallax, and direct ptychography we find that the SSNR has closed-form analytical expressions and is indeed dose-independent (Bennemann et al., 2024, Seki et al., 2018, Varnavides et al., 2023). Surprisingly, for iterative ptychography, we find that the SSNR is dose-dependent: in the low-dose limit, it approaches that of direct ptychography, while at higher doses, it saturates to a value of $\sqrt{2}/2$. We place this observation in the context of recent work on detective quantum efficiency (DQE) (Bennemann et al., 2024, Yu et al., 2024), and quantum Fisher information limits on diffractive imaging (Dwyer and Paganin, 2024, Vega Ibáñez and Verbeeck, 2025).

## Contrast Transfer of Information

Mathematically, the CTF describes the assumed linear relationship between the Fourier sample phase and spatial frequencies present in the reconstructed image, given by:

$$\begin{aligned}\tilde{I}_j(\boldsymbol{q}) &= 2\tilde{\varphi}(\boldsymbol{q}) \times \mathcal{L}_j(\boldsymbol{q}) \\ \tilde{I}_j(\boldsymbol{q}) &= \mathcal{F}_{\boldsymbol{r}\to\boldsymbol{q}}[I_j(\boldsymbol{r})],\end{aligned} \quad (1)$$

where $\tilde{\varphi}(\boldsymbol{q})$ and $\tilde{I}_j(\boldsymbol{q})$ are the Fourier sample phase and reconstructed image, evaluated at spatial frequency $\boldsymbol{q}$.

The complex-valued CTF for the $j^{\text{th}}$ detector segment, $\mathcal{L}_j(\boldsymbol{q})$, is given by (Hammel and Rose, 1995):

$$\mathcal{L}_j(\boldsymbol{q}) = \frac{i}{2}\int D_j(\boldsymbol{k})[\psi^*(\boldsymbol{k})\psi(\boldsymbol{q}-\boldsymbol{k}) - \psi(\boldsymbol{k})\psi^*(\boldsymbol{q}+\boldsymbol{k})], \quad (2)$$

where $\psi(\boldsymbol{k}) = A(\boldsymbol{k})e^{-i\chi(\boldsymbol{k})}$ is the complex-valued probe wavefunction, evaluated at detector frequency, $\boldsymbol{k}$. $A(\boldsymbol{k})$ is the probe-forming aperture given by a normalized top-hat function and $\chi(\boldsymbol{k})$ is the aberration surface given by:

$$\chi(k,\theta) = \frac{2\pi}{\lambda}\sum_{n,m}\frac{1}{n+1}C_{n,m}(k\lambda)^{n+1}\cos[m(\theta - \theta_{n,m})] \quad (3)$$

where $k = |\boldsymbol{k}|$, $\theta = \arctan[\boldsymbol{k}]$, $\lambda$ is the relativistic electron wavelength, $n$ and $m$ are radial and azimuthal orders of the aberration coefficients $C_{n,m}$ with aberration axis $\theta_{n,m}$.

Equation 2 can be expressed more compactly as (Bekkevold et al., 2025):

$$\begin{aligned}\mathfrak{R}\{\mathcal{L}_j(\boldsymbol{q})\} &= \frac{1}{2}\{[\psi \star \psi D_j](\boldsymbol{q}) + [\psi D_j \star \psi](\boldsymbol{q})\} \\ \mathfrak{I}\{\mathcal{L}_j(\boldsymbol{q})\} &= \frac{1}{2}\{[\psi \star \psi D_j](\boldsymbol{q}) - [\psi D_j \star \psi](\boldsymbol{q})\},\end{aligned} \quad (4)$$

where $\star$ denotes cross-correlation and $\mathfrak{R}\{\mathcal{L}\},\mathfrak{I}\{\mathcal{L}\}$ denote the real and imaginary parts of the complex-valued CTF.

Figure 2 demonstrates the components of Equation 4 schematically for a pixelated detector, $D_j(\boldsymbol{k}) = \delta(\boldsymbol{k})$. For an ideal in-focus probe with $\chi(\boldsymbol{q}) = 0$ (Figure 2 a), the CTF is given by the aperture autocorrelation function, which limits all direct STEM phase retrieval techniques. This can be interpreted geometrically as the 'double overlap' area between the aperture and a shifted aperture at a specific spatial frequency (Figure 2 b). In the presence of probe aberrations (Figure 2 c, here showing defocus), the complex probe autocorrelation exhibits negative components, i.e. contrast reversals. This can be understood by noticing that overlaying a shifted and conjugated probe with the complex probe results in so-called 'achromatic lines' (Pennycook et al., 2019), with both positive and negative contributions. Summing these at each spatial frequency recovers the complex probe autocorrelation function.

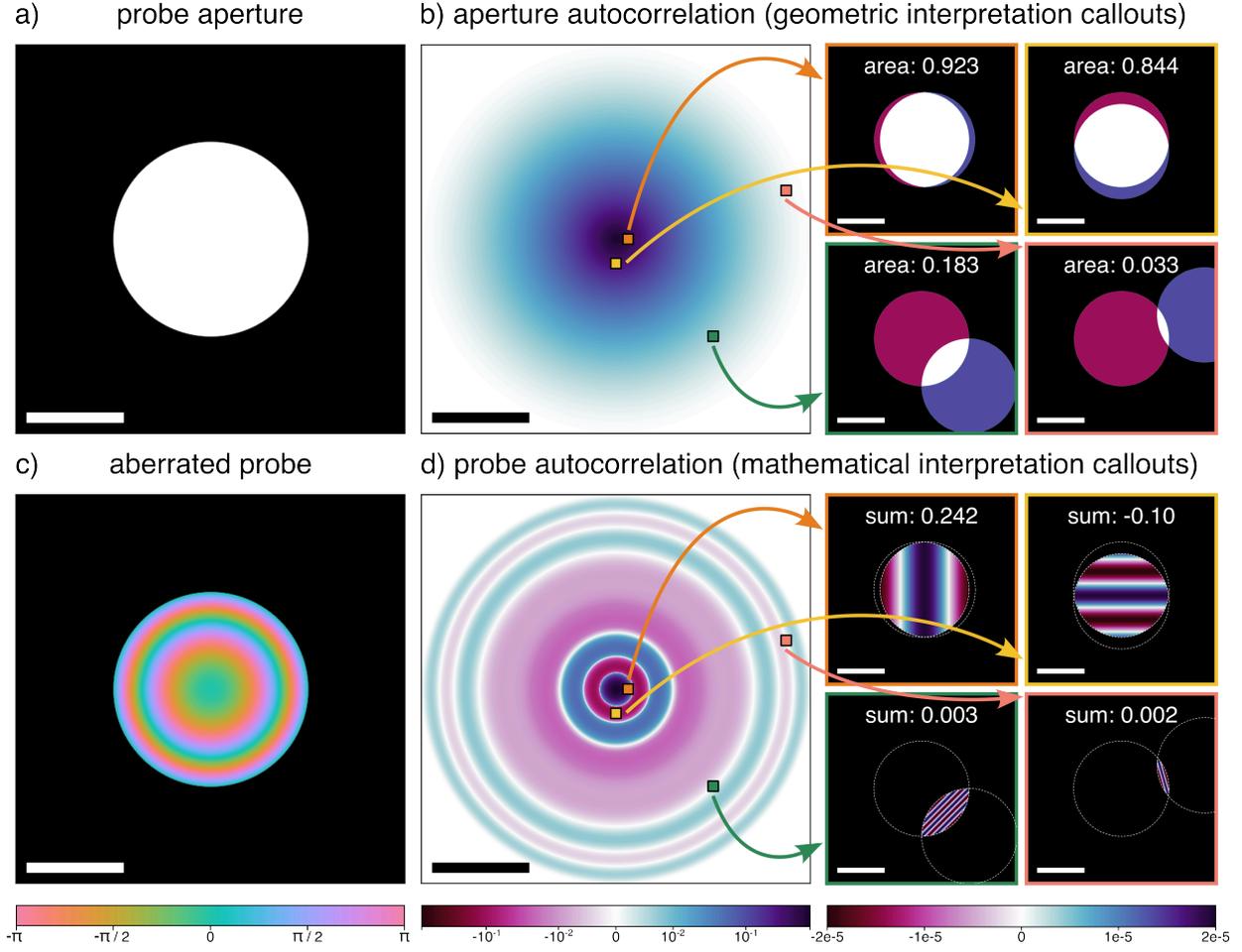

Figure 2: Geometric and mathematical interpretation of the probe autocorrelation functions limiting STEM-based phase retrieval. For an ideal probe, the aperture autocorrelation is understood as the 'double overlap' area for each spatial frequency. Similarly, in the presence of aberrations the 'double overlap' region exhibits positive and negative contributions ('achromatic lines') which are summed to obtain the final autocorrelation value.

## Center of Mass Imaging

The center-of-mass imaging CTF for each Cartesian direction is obtained by Equation 4 by using a vectorial detector function proportional to the detector plane position (Bekkevold et al., 2025, Lazić and Bosch, 2017), $D(k) = k$

$$\mathcal{L}_x^{\text{COM}}(q) = \frac{1}{2}\{[\psi \star \psi\, q_x](q) + [\psi\, q_x \star \psi](q)\}$$
$$\mathcal{L}_y^{\text{COM}}(q) = \frac{1}{2}\{[\psi \star \psi\, q_y](q) - [\psi\, q_y \star \psi](q)\}. \quad (5)$$

The vectorial CTFs in Equation 5 can be Fourier-integrated to provided the scalar center-of-mass imaging CTF

$$\mathcal{L}^{\text{iCOM}}(q) = \frac{q_x \mathcal{L}_x^{\text{COM}}(q) + q_y \mathcal{L}_y^{\text{COM}}(q)}{\text{i}\,|q|^2}, \quad (6)$$

which reduces to the simpler (Bekkevold et al., 2025):

$$\mathcal{L}^{\text{iCOM}} = \frac{\text{i}}{2}[\psi \star \psi](q). \quad (7)$$

Equation 7 is precisely the complex probe autocorrelation function shown in Figure 2.

## Direct Ptychography

The direct ptychography CTF is obtained by recognizing that, even in the presence of aberrations, modern ptychographic algorithms can correct for phase and amplitude variations in this 'double overlap' region (Yang et al., 2016). As such, the direct ptychography CTF is obtained by summing all nonzero contributions in the integrand of Equation 2, known as the aperture-overlap function:

$$\mathcal{L}^{\text{ptycho}}(q) = \frac{\text{i}}{2} \int |\Gamma(q, k)|\, dk$$
$$\Gamma(q, k) = \psi^*(k)\psi(q-k) - \psi(k)\psi^*(q+k). \quad (8)$$

## Parallax Imaging

Formally, parallax imaging can be understood by factoring out a phase-ramp term – representing the apparent lateral shifts – from the aperture-overlap function, $\Gamma(q, k)$, see Bekkevold et al. (2025), Yu et al. (2024). To keep the exposition brief, we simply present the intuitive result that, since the technique amounts to aligning virtual bright field images, the parallax CTF is given by the TEM

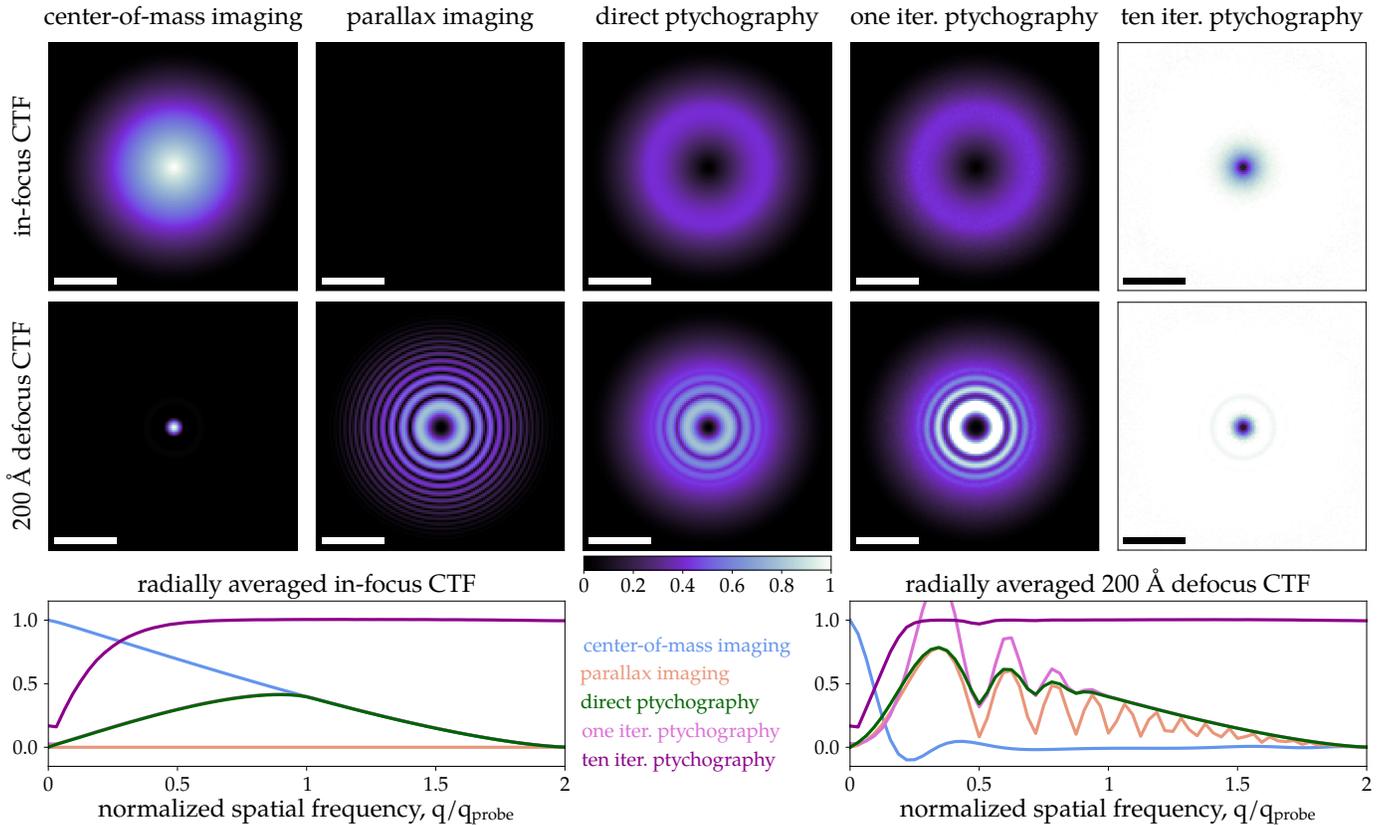

Figure 3: Center-of-mass imaging, parallax imaging, direct ptychography, and iterative ptychography CTFs for acquisitions in-focus and 200 Å defocus, highlighting that center-of-mass and parallax imaging should be performed in-focus and defocused respectively. The direct ptychography CTF is closely related to parallax imaging, albeit recovering the latter's zero crossings and is readily recovered by a single iteration of iterative ptychography. With increasing iterations, the iterative ptychography CTF approaches unity, recovering the low spatial frequencies last.

CTF, modulated by the aperture autocorrelation envelope (Bekkevold et al., 2025, Ma et al., 2025, Yu et al., 2024):

$$\mathcal{L}^{\text{parallax}}(\boldsymbol{q}) = -\frac{i}{2} \sin[\chi(\boldsymbol{q})][A \star A](\boldsymbol{q}). \qquad (9)$$

**Iterative Ptychography**

The iterative nature and sophisticated model-based forward models of iterative ptychography preclude, to our knowledge, an analytical expression for the CTF. However, a number of recent numerical studies have shown that, given sufficient iterations, iterative ptychography can exactly recover all the spatial frequencies in the object – i.e. its CTF is unity, $\mathcal{L}^{\text{iter. ptycho}}(\boldsymbol{q}) = 1$ (Bekkevold et al., 2025, Ma and Muller, 2024, Varnavides et al., 2023).

**CTF Comparison & Limitations**

Figure 3 summarizes the above analytical expressions for in-focus and defocused acquisitions[i]. We make the following observations (Bekkevold et al., 2025):

1. Center-of-mass imaging should be performed in-focus, as its CTF degrades rapidly with probe aberrations, and can in-fact exhibit contrast reversals at large defocus.

2. In contrast, parallax imaging has zero contrast in-focus and requires aberrations to form image contrast. Its CTF exhibits zero-crossings and must be phase-flipped.

3. The direct ptychography CTF successfully recovers the zero-crossings in the parallax CTF and shows strong contrast for both in-focus and defocused acquisitions.

4. A single iteration of iterative ptychography recovers the direct ptychography CTF. Subsequent iterations recover all remaining frequencies, albeit very slowly at low spatial frequencies (McCray et al., 2025).

It's important to note that, the CTFs presented so far can be misleading in evaluating the practical performance of STEM phase retrieval techniques. For example, notice how the center-of-mass imaging CTF approaches unity as the spatial frequency approaches zero (see Figure 3). This is incorrect for all in-line holography methods, such as STEM phase retrieval techniques, since a reference wave is required to recover the absolute phase of a sample. Similarly, as alluded to in the introduction, experimental power spectra of iterative ptychography reconstructions at low dose (see Figure 1), exhibit characteristic 'Thon-like' oscillations (Küçükoğlu et al., 2024), suggesting the iterative ptychography transfer of information is not unity.

---

[i] All numerical results are performed at 300 kV, with a 20 mrad convergence semiangle, and a maximum scattering angle twice that.

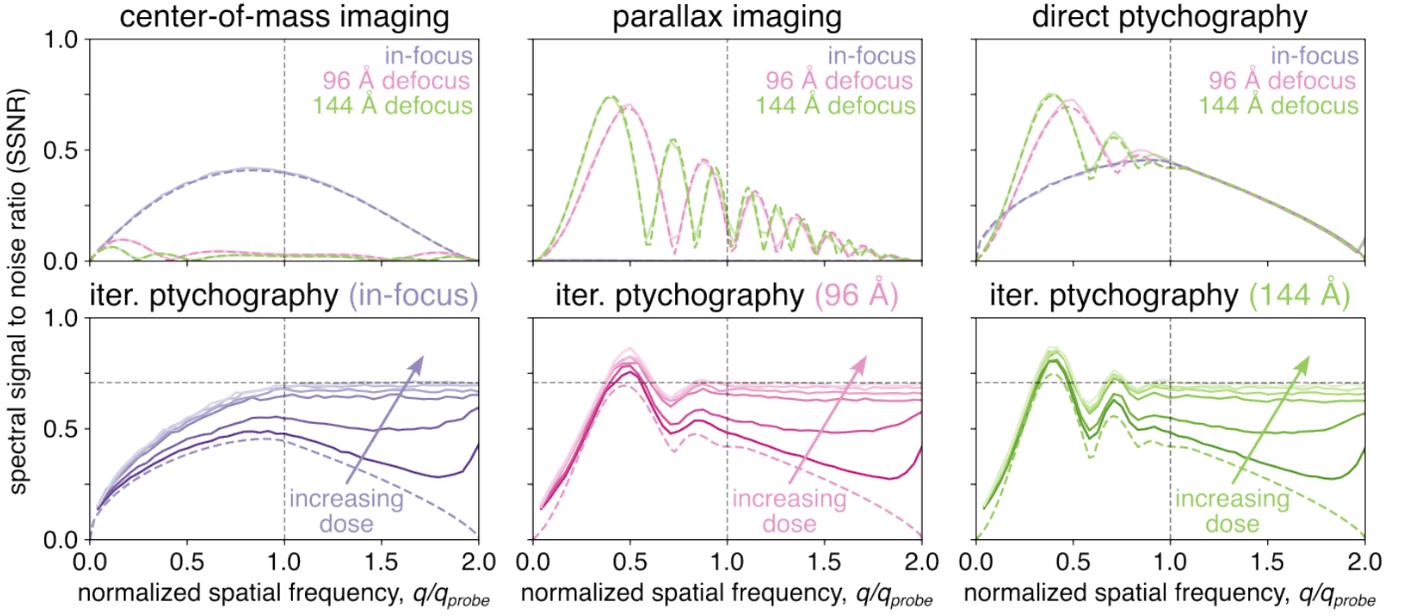

Figure 4: Numerical (solid lines) spectral signal-to-noise ratios for STEM phase retrieval techniques for in-focus and defocused acquisitions. For center-of-mass imaging, parallax imaging, and direct ptychography the numerical SSNRs are dose-independent and well-described by closed-form analytical expressions (dotted lines). By contrast, iterative ptychography exhibits dose-dependent SSNRs, with the low-fluence limit in good agreement with direct ptychography and the high-fluence limit given by recent quantum Fisher information bounds.

## Spectral Signal-to-Noise Ratio

To address the shortcomings of the CTF metric, we turn to a statistical framework, which evaluates the reconstruction fidelity of $M$ numerical reconstructions of the same object, using the SSNR metric (Unser et al., 1987):

$$\text{SSNR}(\boldsymbol{q}) = \frac{|\sum_i \Phi_i(\boldsymbol{q})/M|}{\sqrt{\sum_i |\Phi_i(\boldsymbol{q}) - \overline{\Phi}(\boldsymbol{q})|/(M-1)}} \quad (10)$$
$$\Phi_i(\boldsymbol{q}) = \mathcal{F}_{\boldsymbol{r} \to \boldsymbol{q}}[\varphi_i(\boldsymbol{r})],$$

where the overline denotes an average and $\varphi_i(\boldsymbol{r})$ is the $i^{\text{th}}$ reconstruction. Intuitively, Equation 10 represents the absolute value of the mean Fourier spectrum of the reconstructions, i.e. the CTF, divided by the sample standard deviation for each spatial frequency.

The SSNR metric is closely related to the DQE, which was recently proposed as a quantitative way to evaluate STEM transfer of information (Bennemann et al., 2024):

$$\text{DQE}(\boldsymbol{q}) = \frac{\text{SSNR}_{\text{out}}^2(\boldsymbol{q})}{\text{SSNR}_{\text{in}}^2(\boldsymbol{q})}, \quad (11)$$

where $\text{SSNR}_{\text{in}}(\boldsymbol{q})$ denotes the SSNR of a reference or 'ideal' imaging system, usually taken to be high-resolution TEM with an ideal Zernike phase-plate (Dwyer and Paganin, 2024, Vega Ibáñez and Verbeeck, 2025, Zernike, 1942).

## White Noise Object Reconstructions

In order to isolate the effect of the sample potential from the performance of the imaging system, we reconstruct white-noise objects which have equal scattering information across all spatial frequencies. Note that, for our choice of white-noise objects, $\text{SSNR}_{\text{in}}(\boldsymbol{q}) = \sqrt{N_e}$, where $N_e$ is the total electron dose on the sample, and as such Equation 11 is trivially related to the simpler SSNR metric given by Equation 10, which we focus on in what follows.

Figure 4 plots the SSNR results for $M = 100$ independent reconstructions and three different imaging conditions: in-focus, 96 Å, and 144 Å defocus. For center-of-mass imaging, we see that, as expected, the SSNR metric tends to zero for both $\boldsymbol{q} \to 0$ and $\boldsymbol{q} \to 2q_{\text{probe}}$. For parallax, the SSNR metric is identically equal to its CTF, suggesting the noise in the denominator of Equation 10 is constant. The direct ptychography SSNR is likewise very closely related to its CTF since, as we will see, the noise in the denominator is proportional to the square root of its CTF.

The SSNRs investigated so far were validated across different electron doses and are, as expected, dose-independent. Curiously, this is not the case for iterative ptychography (Figure 4, bottom). At low electron fluence, the SSNR is very similar to that of direct ptychography (dashed lines), suggesting that the algorithm is not able to efficiently extract the necessary information, appearing 'unconverged' (Varnavides et al., 2023). We note that this might be a limitation of the fixed step-size used in the stochastic gradient descent implementation.

At higher electron fluence, iterative ptychography accurately recovers the higher spatial frequency information, with the SSNR saturating at a value of $\sqrt{2}/2$. Recalling the relationship between SSNR and DQE, this implies a DQE limit of 1/2 compared to ideal Zernike phase-contrast – in agreement with the recent theoretical arguments by Dwyer and Paganin (2024). Note that, even at high electron fluence, iterative ptychography still struggles to accurately recover low spatial frequency information.

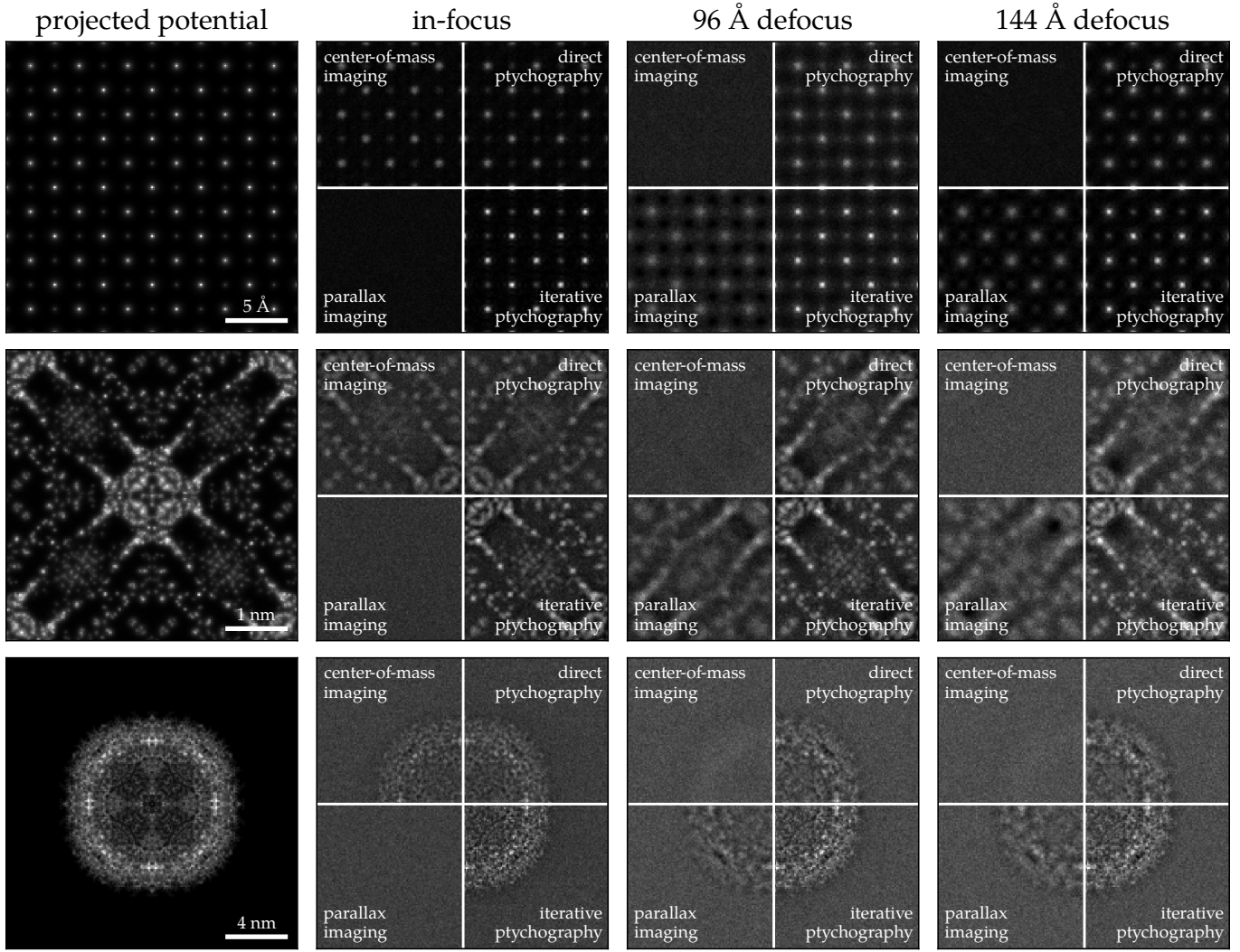

Figure 5: Summary figure convolving relatively low-fluence SSNRs of various STEM phase retrieval techniques for three different imaging conditions with three atomic potentials across different length-scales. Note these are not numerical reconstructions from simulated diffraction intensities, but rather leverage the assumed linearity in Equation 1 to convolve the projected potentials (left column) with the numerical 2D SSNR($q$) shown in Figure 4.

## Derived Analytical Expressions

The dashed lines in the top panels of Figure 4 are analytical closed-form SSNR expressions for center-of-mass imaging, parallax imaging, and direct ptychography. The signal in the SSNR numerator is given by $|\mathcal{L}(q)|$, while the denominator noise has been fitted from Figure 4 to be:

$$\begin{aligned}
\text{Noise}^{\text{iCOM}}(q) &= 1/q \\
\text{Noise}^{\text{parallax}}(q) &= 1 \\
\text{Noise}^{\text{ptycho}}(q) &= \sqrt{\mathcal{L}^{\text{ptycho}}(q)/2}
\end{aligned} \quad (12)$$

Equation 12 highlights how the noise in center-of-mass imaging is inversely proportional to the spatial frequency. This well-known result (Varnavides et al., 2023, Yu et al., 2024), resolves the apparent paradox in the center-of-mass imaging CTF being equal to 1 at zero spatial frequency. The noise in parallax imaging being constant can be understood by recalling that the implicit operation parallax performs – virtual bright field alignment – can be aided by higher contrast fiducial markers. Similarly, the noise of direct ptychography being proportional to the square root of its CTF is motivation for the normalization kernel used in the related optimum bright-field (OBF) STEM technique (Ooe et al., 2021, 2024). The analytical parallax imaging and direct ptychography SSNR($q$) expressions are overlaid on the experimental power spectra of biological crystals shown in Figure 1, to good agreement – highlighting that indeed at low fluence the iterative ptychography SSNR($q$) approaches that of direct ptychography.

## Conclusions

Figure 5 summarizes the SSNR results by convolving the projected atomic potentials of various samples across length-scales with the 2D SSNR($q$) presented throughout this work, highlighting the following key observations:

1. Center-of-mass imaging should be performed in-focus to maximize information and avoid contrast reversals.
2. Parallax imaging should be performed out of focus to obtain contrast, albeit at the expense of zero-crossings.
3. At low electron fluence, direct and iterative ptychography exhibit only minor differences at high frequencies.

These findings reinforce that dose efficiency and information transfer in STEM phase retrieval are not fully captured by conventional CTF analysis (Bennemann et al., 2024, Seki et al., 2018, Varnavides et al., 2023, Yu et al., 2024). By leveraging SSNR, a dose-aware statistical metric, we provide a robust framework for evaluating the practical performance of STEM phase retrieval techniques.

In particular, we demonstrate that while the CTF of iterative ptychography appears ideal in theory, its recoverable signal content can fall short in low-dose regimes. Conversely, we find that more computationally-efficient direct imaging techniques, such as parallax imaging and direct ptychography, may be comparably effective for dose-sensitive samples. We anticipate the SSNR, or equivalently DQE, metric will serve as a valuable tool for guiding the development and optimization of phase retrieval acquisitions at low electron fluence.

## Availability of Data and Materials

The experimental dataset on biological crystals are obtained from Küçükoğlu et al. (2024). All other processed datasets/notebooks are available at [Add Zenodo link].

## Author Contributions Statement

G.V.: conceptualization, data curation, formal analysis, investigation, methodology, software, validation, visualization, writing – original draft, writing – review & editing; J.M.B., S.M.R.: investigation, validation, visualization, writing - review & editing; M.C.S, L.J.: funding acquisition, project administration, supervision, writing - review & editing; C.O.: conceptualization, formal analysis, funding acquisition, investigation, methodology, project administration, software, supervision, writing – review & editing; All authors read and approved the final version of the manuscript.

## Acknowledgements

The authors gratefully acknowledge Berk Küçükoğlu and Prof. Henning Stahlberg for providing the experimental dataset on biological crystals used to motivate this study. The authors gratefully acknowledge Profs. David A. Muller and Peter Nellist for insightful discussions and for drawing their attention to the recent results on Quantum Fisher information.

## Financial Support

Work at the Molecular Foundry was supported by the Office of Science, Office of Basic Energy Sciences, of the U.S. Department of Energy under Contract No. DE-AC02-05CH11231. J.M.B. and L.J. acknowledge support from Research Ireland grant number 19/FFP/6813. L.J. acknowledges support from Royal Society and Research Ireland grant numbers URF/RI/191637 and 12/RC/2278_P2. S.M.R. and C.O. acknowledge support from the DOE Early Career Research Program.

## Conflict of Interest

The authors declare there are no conflicts of interest.

## Bibliography

Bekkevold, J.M., Peters, J.J.P., Ishikawa, R., Shibata, N., Jones, L., 2024. Ultra-fast Digital DPC Yielding High Spatio-temporal Resolution for Low-Dose Phase Characterization. Microscopy and Microanalysis 30, 878–888.. https://doi.org/10.1093/mam/ozae082

Bekkevold, Julie Marie, Ribet, S.M., Scott, M.C., Jones, L., Ophus, Colin, Varnavides, G., 2025. Evaluating the Transfer of Information in Phase Retrieval STEM Techniques. Elemental Microscopy.. https://doi.org/10.69761/ehch7395

Bennemann, F., Nellist, P., Kirkland, A., 2024. Quantitative comparison of HRTEM and electron ptychography. BIO Web of Conferences 129, 4007.. https://doi.org/10.1051/bioconf/202412904007

Chen, Z., Odstrcil, M., Jiang, Y., Han, Y., Chiu, M.-H., Li, L.-J., Muller, D.A., 2020. Mixed-state electron ptychography enables sub-angstrom resolution imaging with picometer precision at low dose. Nature Communications 11.. https://doi.org/10.1038/s41467-020-16688-6

Dekkers, N.H., De Lang, H., 1974. Differential phase contrast in a STEM. Optik 41, 452–456.

Dwyer, C., Paganin, D.M., 2024. Quantum and classical Fisher information in four-dimensional scanning transmission electron microscopy. Physical Review B 110.. https://doi.org/10.1103/physrevb.110.024110

Egerton, R., 2013. Control of radiation damage in the TEM. Ultramicroscopy 127, 100–108.. https://doi.org/10.1016/j.ultramic.2012.07.006

Egerton, R., 2007. Limits to the spatial, energy and momentum resolution of electron energy-loss spectroscopy. Ultramicroscopy 107, 575–586.. https://doi.org/10.1016/j.ultramic.2006.11.005

Hammel, M., Rose, H., 1995. Optimum rotationally symmetric detector configurations for phase-contrast imaging in scanning transmission electron microscopy. Ultramicroscopy 58, 403–415.. https://doi.org/10.1016/0304-3991(95)00007-n

Küçükoğlu, B., Mohammed, I., Guerrero-Ferreira, R.C., Ribet, S.M., Varnavides, G., Leidl, M.L., Lau, K., Nazarov, S., Myasnikov, A., Kube, M., Radecke, J., Sachse, C., Müller-Caspary, K., Ophus, C., Stahlberg, H., 2024. Low-dose cryo-electron ptychography of proteins at sub-nanometer resolution. Nature Communications 15.. https://doi.org/10.1038/s41467-024-52403-5

Lazić, I., Bosch, E.G., 2017. Analytical Review of Direct Stem Imaging Techniques for Thin Samples, in: Advances in Imaging and Electron Physics. Elsevier, pp. 75–184.. https://doi.org/10.1016/bs.aiep.2017.01.006

Lazić, I., Bosch, E.G., Lazar, S., 2016. Phase contrast STEM for thin samples: Integrated differential phase contrast. Ultramicroscopy 160, 265–280.. https://doi.org/10.1016/j.ultramic.2015.10.011

Li, G., Xu, M., Tang, W.-Q., Liu, Y., Chen, C., Zhang, D., Liu, L., Ning, S., Zhang, H., Gu, Z.-Y., Lai, Z., Muller, D.A., Han, Y., 2025. Atomically resolved imaging of radiation-sensitive metal-organic frameworks via electron ptychography. Nature Communications 16.. https://doi.org/10.1038/s41467-025-56215-z

Liu, L., Zhang, D., Zhu, Y., Han, Y., 2020. Bulk and local structures of metal–organic frameworks unravelled by high-resolution electron microscopy. Communications Chemistry 3.. https://doi.org/10.1038/s42004-020-00361-6

Ma, D., Li, G., Muller, D.A., Zeltmann, S.E., 2025. Information in 4D-STEM: Where it is, and How to Use it [WWW Document].. https://doi.org/10.48550/ARXIV.2507.21034

Ma, D., Muller, D., 2024. Information Limit and Dose Efficiency of Electron Ptychography. Microscopy and Microanalysis 30.. https://doi.org/10.1093/mam/ozae044.910

McCray, A.R.C., Ribet, S.M., Varnavides, G., Ophus, C., 2025. Accelerating iterative ptychography with an integrated neural network. Journal of Microscopy.. https://doi.org/10.1111/jmi.13407

Ooe, K., Seki, T., Ikuhara, Y., Shibata, N., 2021. Ultra-high contrast STEM imaging for segmented/pixelated detectors by maximizing the signal-to-noise ratio. Ultramicroscopy 220, 113133.. https://doi.org/10.1016/j.ultramic.2020.113133

Ooe, K., Seki, T., Nogami, M., Ikuhara, Y., Shibata, N., 2024. Dose-efficient phase-contrast imaging of thick weak phase objects via OBF STEM using a pixelated detector. Microscopy 74, 98–106.. https://doi.org/10.1093/jmicro/dfae051

O'Leary, C.M., Martinez, G.T., Liberti, E., Humphry, M.J., Kirkland, A.I., Nellist, P.D., 2021. Contrast transfer and noise considerations in focused-probe electron ptychography. Ultramicroscopy 221, 113189.. https://doi.org/10.1016/j.ultramic.2020.113189

Pennycook, T.J., Lupini, A.R., Yang, H., Murfitt, M.F., Jones, L., Nellist, P.D., 2015. Efficient phase contrast imaging in STEM using a pixelated detector. Part 1: Experimental demonstration at atomic resolution. Ultramicroscopy 151, 160–167.. https://doi.org/10.1016/j.ultramic.2014.09.013

Pennycook, T.J., Martinez, G.T., Nellist, P.D., Meyer, J.C., 2019. High dose efficiency atomic resolution imaging via electron ptychography. Ultramicroscopy 196, 131–135.. https://doi.org/10.1016/j.ultramic.2018.10.005

Rodenburg, J., Maiden, A., 2019. Ptychography, in: Springer Handbook of Microscopy. Springer International Publishing, pp. 819–904.. https://doi.org/10.1007/978-3-030-00069-1_17

Seki, T., Ikuhara, Y., Shibata, N., 2018. Theoretical framework of statistical noise in scanning transmission electron microscopy. Ultramicroscopy 193, 118–125.. https://doi.org/10.1016/j.ultramic.2018.06.014

Spoth, K.A., Nguyen, K.X., Muller, D.A., Kourkoutis, L.F., 2017. Dose-Efficient Cryo-STEM Imaging of Whole Cells Using the Electron Microscope Pixel Array Detector. Microscopy and Microanalysis 23, 804–805.. https://doi.org/10.1017/s1431927617004688

Unser, M., Trus, B.L., Steven, A.C., 1987. A new resolution criterion based on spectral signal-to-noise ratios.

Ultramicroscopy 23, 39–51.. https://doi.org/[10.1016/0304-3991(87)90225-7](10.1016/0304-3991(87)90225-7)

Varnavides, G., Ribet, S.M., Ophus, C., 2024. Tilt-Corrected BF-STEM. Elemental Microscopy.. https://doi.org/[10.69761/xunr2166](10.69761/xunr2166)

Varnavides, G., Ribet, S.M., Zeltmann, S.E., Yu, Y., Savitzky, B.H., Byrne, D.O., Allen, F.I., Dravid, V.P., Scott, M.C., Ophus, C., 2023. Iterative Phase Retrieval Algorithms for Scanning Transmission Electron Microscopy [WWW Document].. https://doi.org/[10.48550/ARXIV.2309.05250](10.48550/ARXIV.2309.05250)

Vega Ibáñez, F., Verbeeck, J., 2025. Retrieval of Phase Information from Low-Dose Electron Microscopy Experiments: Are We at the Limit Yet?. Microscopy and Microanalysis 31.. https://doi.org/[10.1093/mam/ozae125](10.1093/mam/ozae125)

Yang, H., Ercius, P., Nellist, P.D., Ophus, C., 2016. Enhanced phase contrast transfer using ptychography combined with a pre-specimen phase plate in a scanning transmission electron microscope. Ultramicroscopy 171, 117–125.. https://doi.org/[10.1016/j.ultramic.2016.09.002](10.1016/j.ultramic.2016.09.002)

Yang, H., Pennycook, T.J., Nellist, P.D., 2015. Efficient phase contrast imaging in STEM using a pixelated detector. Part II: Optimisation of imaging conditions. Ultramicroscopy 151, 232–239.. https://doi.org/[10.1016/j.ultramic.2014.10.013](10.1016/j.ultramic.2014.10.013)

Yu, Y., Spoth, K.A., Colletta, M., Nguyen, K.X., Zeltmann, S.E., Zhang, X.S., Paraan, M., Kopylov, M., Dubbeldam, C., Serwas, D., Siems, H., Muller, D.A., Kourkoutis, L.F., 2024. Dose-Efficient Cryo-Electron Microscopy for Thick Samples using Tilt-Corrected Scanning Transmission Electron Microscopy, Demonstrated on Cells and Single Particles.. https://doi.org/[10.1101/2024.04.22.590491](10.1101/2024.04.22.590491)

Zernike, F., 1942. Phase contrast, a new method for the microscopic observation of transparent objects part II. Physica 9, 974–986.. https://doi.org/[10.1016/s0031-8914(42)80079-8](10.1016/s0031-8914(42)80079-8)